\documentclass[%
 reprint,
 amsmath,amssymb,
 aps,
prb,
]{revtex4-2}

\usepackage[
]{geometry}

\usepackage{graphicx}
\usepackage{dcolumn}
\usepackage{bm}
\usepackage{xcolor}
\usepackage[normalem]{ulem}
\usepackage[hidelinks]{hyperref}
\usepackage[utf8]{inputenc}
\usepackage[T1]{fontenc}
\usepackage{mathptmx}
\usepackage{etoolbox}
\usepackage{cleveref}
\usepackage{lmodern}

\begin{document}
\newcommand\redsout{\bgroup\markoverwith{\textcolor{red}{\rule[0.5ex]{2pt}{0.4pt}}}\ULon}

\renewcommand{\sec}[1]{\section{#1}}
\newcommand{\ssec}[1]{\subsection{#1}}
\newcommand{\sssec}[1]{\subsubsection{#1}}


\def\bea{\begin{eqnarray}}
\def\eea{\end{eqnarray}}
\def\ben{\begin{equation}}
\def\een{\end{equation}}
\def\benu{\begin{enumerate}}
\def\enu{\end{enumerate}}

\def\bei{\begin{itemize}}
\def\eei{\end{itemize}}
\def\beit{\begin{itemize}}
\def\eit{\end{itemize}}
\def\benu{\begin{enumerate}}
\def\enu{\end{enumerate}}

\def\n{n}
\def\np{{n^{\prime}}}
\def\npp{{n^{\prime \prime}}}

\def\sss{\scriptscriptstyle\rm}

\def\g{_\gamma}

\def\l{^\lambda}
\def\lfc{^{\lambda=1}}
\def\lo{^{\lambda=0}}

\def\marnote#1{\marginpar{\tiny #1}}
\def\rsav{\langle r_s \rangle}
\def\invdif{\frac{1}{|\br_1 - \br_2|}}

\def\hatT{{\hat T}}
\def\hatV{{\hat V}}
\def\hatH{{\hat H}}
\def\1var{(\bx_1...\bx\N)}

\def\half{\frac{1}{2}}
\def\quart{\frac{1}{4}}

\def\bp{{\bf p}}
\def\br{{\bf r}}
\def\bR{{\bf R}}
\def\bu{{\bf u}}
\def\bx{{x}}
\def\by{{y}}
\def\ba{{\bf a}}
\def\bq{{\bf q}}
\def\bj{{\bf j}}
\def\bX{{\bf X}}
\def\bF{{\bf F}}
\def\bchi{{\bf \chi}}
\def\bof{{\bf f}}

\def\cA{{\cal A}}
\def\cB{{\cal B}}

\def\x{_{\sss X}}
\def\c{_{\sss C}}
\def\s{_{\sss S}}
\def\xc{_{\sss XC}}
\def\Hx{_{\sss HX}}
\def\Hxc{_{\sss HXC}}
\def\xj{_{{\sss X},j}}
\def\xcj{_{{\sss XC},j}}
\def\N{_{\sss N}}
\def\H{_{\sss H}}

\def\t{^{\tau}}

\def\ext{_{\rm ext}}
\def\pot{^{\rm pot}}
\def\hyb{^{\rm hyb}}
\def\HF{^{\rm HF}}
\def\hah{^{1/2\& 1/2}}
\def\loc{^{\rm loc}}
\def\LSD{^{\rm LSD}}
\def\LDA{^{\rm LDA}}
\def\GEA{^{\rm GEA}}
\def\GGA{^{\rm GGA}}
\def\SPL{^{\rm SPL}}
\def\sce{^{\rm SCE}}
\def\PBE{^{\rm PBE}}
\def\DFA{^{\rm DFA}}
\def\TF{^{\rm TF}}
\def\VW{^{\rm VW}}
\def\helm{^{\rm unamb}}
\def\una{^{\rm unamb}}
\def\ion{^{\rm ion}}
\def\HOMO{^{\rm HOMO}}
\def\LUMO{^{\rm LUMO}}
\def\gs{^{\rm gs}}
\def\dyn{^{\rm dyn}}
\def\adia{^{\rm adia}}
\def\I{^{\rm I}}
\def\pot{^{\rm pot}}
\def\sav{^{\rm sph. av.}}
\def\syv{^{\rm sys. av.}}
\def\pnav{^{\rm sym}}
\def\av#1{\langle #1 \rangle}
\def\unif{^{\rm unif}}
\def\LSD{^{\rm LSD}}
\def\ee{_{\rm ee}}
\def\vir{^{\rm vir}}
\def\ALDA{^{\rm ALDA}}
\def\PGG{^{\rm PGG}}
\def\GK{^{\rm GK}}
\def\atom{^{\rm atmiz}}
\def\trans{^{\rm trans}}
\def\unpol{^{\rm unpol}}
\def\pol{^{\rm pol}}
\def\sav{^{\rm sph. av.}}
\def\syv{^{\rm sys. av.}}

\def\up{_\uparrow}
\def\dn{_\downarrow}
\def\upp{\uparrow}
\def\dnn{\downarrow}

\def\shalf{$^{\hspace{0.006cm}1}\! \! \hspace{0.045cm} / _{\! 2}$}

\def\td{time-dependent~}
\def\KS{Kohn-Sham~}
\def\DFT{density functional theory~}

\def\fourint{ \int_{t_0}^{t_1} \! dt \int \! d^3r\ }
\def\fourintp{ \int_{t_0}^{t_1} \! dt' \int \! d^3r'\ }
\def\intx{\int\!d^4x}
\def\sph_int{ {\int d^3 r}}
\def\radint{ \int_0^\infty dr\ 4\pi r^2\ }
\def\intrrp{\int d^3r \int d^3r'\,}
\def\intr{\int d^3r\,}
\def\intrp{\int d^3r'\,}


\title{Two-legged approximation for building non-empirical hybrids and analyzing correlation at finite temperature}

\author{Brittany P. Harding}
\thanks{These two authors contributed equally.}
\address{University of California, Merced, 5200 North Lake Road, Merced, CA 95343, USA}

\author{Francisca Sagredo}
\thanks{These two authors contributed equally.}
\address{Department of Physics, University of California, Berkeley, California 94720, USA}
\address{Materials Sciences Division, Lawrence Berkeley National Laboratory, Berkeley, California 94720, USA}

\author{Vincent Martinetto}
\address{University of California, Merced, 5200 North Lake Road, Merced, CA 95343, USA}
\author{Aurora Pribram-Jones}
\email{apj@ucmerced.edu}
\address{University of California, Merced, 5200 North Lake Road, Merced, CA 95343, USA}

\date{\today}

\begin{abstract}
Warm dense matter is a highly energetic phase characterized by strong correlations, thermal effects, and quantum effects of electrons. Thermal density functional theory is commonly used in simulations of this challenging phase, driving the development of temperature-dependent approximations to the exchange-correlation free energy. In this work, a finite-temperature extension of the two-legged adiabatic connection construction is demonstrated for the uniform electron gas and asymmetric Hubbard dimer at warm dense matter conditions. This provides the structure of a temperature- and density-dependent weighting scheme for a hybrid exchange-correlation approximation. The construction also provides evidence that nonlinear thermal effects on the balance between exchange and correlation, as well as that between kinetic, entropic, and potential components of the correlation, persist and can even be emphasized by strong electron-electron interaction. These findings point additionally to a complicated interplay between temperature, density, and interaction strength in the strong correlation character of these model systems.
\end{abstract}

\maketitle

\sec{Introduction}

Electrons are the drivers of chemical reactivity, critical in the development of functional materials, and hold the key to progress in quantum computing, energy conversion, and biophysical processes.  Density functional theory (DFT)\cite{HK64,KS65} provides a useful balance of accuracy and computational efficiency and remains a powerful tool for simulations of electronic structure. Though it is an exact theory in principle, its most common implementations rely on carefully crafted approximations to a key piece of the electronic energy, the exchange-correlation (XC) energy. This tiny-but-mighty energy component and its approximations are key to DFT's successes and to many of its failures as well. The accuracy of XC approximations is paramount in simulations of chemical bonding, bond breaking, calculated band gaps, and many other computations for which DFT is still the most accessible choice, even when it sometimes falls short in these challenging areas of electronic structure. 

Some of the most widely used\cite{PGB14} ZT XC functional approximations are hybrid functionals \cite{B93,LYP88,VWN80}, which approximate the exchange component of the XC energy via a combination of Hartree-Fock exchange and that calculated using a DFT method (often a semilocal method for a balance of useful accuracy and computational efficiency). Such approximations have substantially increased the accuracy of the simulations of molecules\cite{} and solids\cite{}. These approaches use various strategies to set the balance of the exact/HF exchange and the GGA exchange, whether via perturbation theory\cite{PEB96} or more direct adiabatic connection approaches,\cite{AB99} empirical fitting to minimize errors on a cleverly selected test set,\cite{B88} or semiclassical reasoning.\cite{EB09} The two-legged representation of the exact adiabatic connection integrand was presented\cite{BEP97} as a non-empirical alternative that generated a density-dependent mixing parameter. Though based on the adiabatic connection, this parameter could be calculated from only three values for any system via the generalized gradient approximation and was designed to improve calculations for approximations that treated combined exchange-correlation energies more accurately than exchange energies.

Though the improvement of approximate XC energy functionals at zero temperature is key to the historic and continued development of DFT, extension of the theory to non-zero temperatures demands that we consider not only how systems vary with density, but also how they depend on temperature. This explicit temperature dependence is especially critical in the study of warm dense matter (WDM). WDM is a highly energetic phase that exists within the interiors of giant planets and in the atmospheres of white dwarf stars \cite{KDLM12,MHVT08,KRDM08}. WDM is also generated experimentally on the path to fusion ignition at notable facilities including Lawrence Livermore National Laboratory’s NIF and Sandia National Laboratory's Z Machine \cite{HCCD16,KD17,MSAA05}. Thermal density functional theory is commonly used to model this challenging phase and is considered the best practice for predictive WDM and high-energy density physics (HEDP) calculations, \cite{WFGG17,KDBL15,RR14,GDRT14,KD09,M65,SP88,VCW14,H17} as it is used to drive ab initio molecular dynamics (AIMD). In order to improve the accuracy of such simulations, researchers have developed FT density functional approximations\cite{KSDT14,KDT18,KMH22}, including a temperature-dependent hybrid approximation.\cite{MKH20} This hybrid approximation was designed to improve FT GGA band gap calculations and did so using a mixing parameter of $1/4$, as justified at ZT in Ref. \cite{PEB96}.

In this work, we first extend the zero-tempearture TLA framework to the finite-temperature adiabatic connection, establishing a formal pathway for a hybrid approximation to the exchange-correlation free energy whose construction has a mixing parameter that is both density- and temperature-dependent. Secondly, we demonstrate that the approach generates exact XC free energies for two exactly solvable models, the FT uniform electron gas and the FT asymmetric Hubbard dimer. Third, we use the temperature- and density-dependent mixing parameter as a proxy for analysis of the relationship between exchange and exchange-correlation of finite-temperature systems, as well as the relationship between kinetic, entropic, and potential components of the correlation free energy. This provides insight into the interplay of temperature, density, and interaction strength that are otherwise inaccessible. Lastly, we discuss how this new framework hints at how the same analysis can yield insights into the exact and approximate FT exchange and exchange-correlation holes in future studies.

\sec{Background}
\ssec{Kohn-Sham DFT}

Hohenberg and Kohn provided the basis for DFT,\cite{HK64} showing that if the exact ground-state density of a many-body interacting system is known, then the ground-state energy can be determined exactly:
    \ben\label{HK}
        E[n({\bf r})] \equiv \min_n \left\{\int d{\bf r}~ v({\bf r})n({\bf r}) + F[n({\bf r})]\right\},
    \een
where $v(\bf r)$ is the external potential, or system-dependent piece, and $F[n(\bf r)]$ is the universal functional, or system-independent piece consisting of the kinetic and electron-electron interaction energies:
    \ben\label{eq:uni}
        F[n(\br)] = T[n(\br)] + V\ee[n(\br)].
    \een

Kohn and Sham provided the framework for the practical implementation of DFT.\cite{KS65} The KS scheme begins with an imaginary system of non-interacting electrons that have the same density as the interacting problem. The electrons experience a potential $v\s$ chosen somehow to imitate the true interacting system, and if we can get the non-interacting system to accurately imitate the physical system, then we will have a much more cost-effective problem at hand. Since the electrons are non-interacting, the coordinates decouple and we may write the wavefunction as a product of single-particle orbitals satisfying
    \ben
        \left\{-\frac{1}{2}\nabla^2+v_s(\br)\right\}\phi_i(\br)=\varepsilon_i\phi_i(\br).
    \een
where $\varepsilon_i$ are the KS eigenvalues, $\phi_i$ are the corresponding KS orbitals, which yield the electronic probability density,
    \ben
        n(\br) = \sum_{i=1}^N |\phi_i(\br)|^2,
    \een
and $v\s$ is the KS potential which is unique by the HK theorem.
The KS wavefunction of orbitals $\Phi=\phi_1\phi_2\dots$ is not considered an approximation to the true wavefunction but is a fundamental property of any electronic system uniquely determined by the electronic density\cite{KS65,ABC}. The KS wavefunction minimizes the kinetic energy to give the kinetic energy of the non-interacting electrons:
    \ben
        T\s[n]=\min_{\Phi\rightarrow n} \langle \Phi|\hatT|\Phi\rangle.
    \een
The $s$ subscript will be used throughout this work to indicate non-interacting quantities. In terms of the non-interacting kinetic energy, Eqn. (\ref{eq:uni}) becomes
    \ben
        F[n(\br)] = T\s[n(\br)] + U\H[n(\br)] + E\xc[n(\br)],
    \een
where $U\H$ is the classical electrostatic repulsion, or Hartree energy, and $E\xc$ is the exchange-correlation (XC) energy, which captures the remainder of interactions not captured by the Hartree energy. $E\xc$ can be broken into an exchange component, $E\x$, and a correlation component, $E\c$, which is further composed of the kinetic and potential correlation components. The kinetic correlation is written 
    \ben
        T\c[n(\br)] = T[n(\br)] - T\s[n(\br)],
    \een
or as the difference between the exact kinetic energy and the non-interacting kinetic energy. The remainder of the correlation energy is the potential correlation, $U\c=E\c-T\c$.

\KS DFT is exact, as long as we have the exact exchange-correlation functional for arbitrary physical systems of interest. Unfortunately, this is not the case for a vast majority of systems, and approximations must be employed for practical calculations.

\ssec{Adiabatic connection formalism}
\label{sec:AC}
Based on teh fluctuation-dissipation theorem, the adiabatic connection\cite{HJ74,LP75,GL76} gives an exact expression for the ground-state exchange-correlation energy, in terms of an integral over the coupling constant, a parameter that alters the electron-electron interaction strength in the electronic Hamiltonian. Integration over this coupling constant $\lambda$, from $\lambda=0$ to $\lambda=1$, smoothly connects the fictitious non-interacting \KS reference system at $\lambda=0$ to the physical interacting system of interest at $\lambda=1$, as long as the density is held fixed throughout the integration process. 

To begin, the coupling constant $\lambda$, is introduced into the universal functional \cite{ABC,LP85}, 

\begin{equation}
F^{\lambda}[n] = \min_{\Psi\to n} \langle \Psi|\hat{T} + \lambda \hat{V}_{ee} |\Psi \rangle,
\end{equation}

\noindent where $\lambda=0$ yields the fictitious KS system and $\lambda=1$ gives the real, interacting system of interest, and at all values of $\lambda$, the density is that of the physical system. The zero-temperature (ZT) adiabatic connection gives an exact expression for the exchange-correlation functional, $E_{xc}[n]$,

\begin{equation}
    E_{xc}[n]=\int_0^1 d\lambda W\xc^{\lambda}[n],
\end{equation}

The integrand, $W_{\lambda}$, is derived via the application of the Hellman-Feynman theorem and is given by $W\xc^\lambda[n]=U\xc^\lambda[n]/\lambda$. 

The finite-temperature adiabatic connection formula (FTACF)\cite{PPFS11,PPGB14} introduces temperature dependence into the traditional AC integrand:
\begin{equation}
    A_{xc}[n]=\int_0^1 d\lambda W\xc^{\tau,\lambda}[n],
\end{equation}
\noindent where it now expresses the XC free energy in terms of the interaction strength-- and temperature-dependent potential XC alone. This formalism was used to analyze an accurate, fully ab initio parametrization\cite{GDSM17} of the exchange-correlation free energy per electron at WDM conditions.\cite{HMP22}

\ssec{Zero-temperature two-legged construction}
Though the ACF is a key interpretive tool and pathway for calculating the exact $E\xc$, we generally do not have easy access to the exact ACF integrand for realistic systems. The two-legged construction of an approximate AC curve was introduced at ZT\cite{BEP97} as a way to benefit from the ACF framework for hybrid XC DFA constructions, despite the ACF's general inaccessiblity. The adiabatic connection can be approximated by two linear legs, the first connecting the energy at $\lambda = 0$ to the energy at $\lambda = b$, and the second connecting the energy at $\lambda = b$ to the energy at $\lambda = 1$. The point $b$ at which the two legs meet is calculated as follows: 
    \ben \label{eq:b}
        b = \frac{E\xc - W\xc^{\lambda=1}}{E\x - W\xc^{\lambda=1}}.
    \een
It is easy to see where this point comes from if one draws an opposite diagonal line that passes through the point $b$, starting at $\lambda = 1$ and ending at $W\xc^{\lambda=1}$, and sums up the areas of the resulting geometric components. The two legs should meet at a point $b$ that results in a two-legged approximation (TLA) with an area that matches the area beneath the exact adiabatic connection curve for a given density. If the point $b$ is restricted to be somewhere along the opposite diagonal, then the area beneath the TLA is the energy at which the two legs meet, i.e., the energy at $\lambda = b$ is the XC energy.

While the adiabatic connection curve appears  almost linear for many systems of interest, it has always been found to be concave upward, i.e.,
    \ben \label{eq:concavity}
        \frac{d^2U\l_c[n]/\lambda}{d\lambda^2} \ge 0,
    \een
although this has not yet been proven rigorously.\cite{PDV24} If $W\xc\l$ is always concave upward, then geometric interpretation implies that $T_c \leq |E_c|$. When written in terms of $T_c$, the parameter $b$ can be interpreted as a measure of dynamic correlation, or the fraction of correlation that is kinetic:
    \ben
        b = \frac{T_c}{|U\c|}.
    \een
This is equivalent to Eqn.(\ref{eq:b}). Due to the condition of Eqn. (\ref{eq:concavity}), $0 < b\leq 1/2$, and $b$ is usually near $0.5$ for real systems. This is because most systems are not far from the high-density limit $r_s\rightarrow \infty$.\cite{} As one moves to lower and lower densities, the value $b$ becomes smaller, coming close to $0$ and only eraching it in the low-density limit. This reflects the dominance of correlation over exchange contributions at low densities. The connection between the kinetic-to-potential ratio and estimates of relative amounts of static and dynamic correlation is rooted in the tendency for weaker scaling of static correlation with density than either $E\x$ or $T\c$. This leads to an expectation of dominance of $T\c$ in the high-density limit and dominance of $U\c$ in the low-density limit. In this way, analysis of the $b$ parameter can give us a window into how a system's proportions of $E\x$, $E\c$, $T\c$, and $U\c$ vary within different density regimes.

To adapt the two-legged construction to useful XC density functional approximations, one can use the $b$ parameter to mix accurate exchange with approximate exchange in a hybrid formulation.
Generalized gradient approximations (GGAs)\cite{PCVJ92,PBW96,B88,LYP88} are popular for calculating chemical reaction energies due to their combined computational efficiency and accuracy\cite{JGP93}. GGAs typically perform better for exchange-correlation energies than for exchange alone, which has led to improvement of calculations by mixing fractions of exact exchange and GGA exchange\cite{B93,B94}. The TLA provides a framework for constructing non-empirical hybrids of GGA energies with exact exchange, where the typical mixing parameter now becomes density-dependent via Eqn. \ref{eq:b} and requires no empirical fitting:
\ben
    E\xc^{\rm hybrid}=b E\x + (1-b) E\xc^{\rm GGA}.
\een
This ZT approach was shown\cite{BEP97} to improve atomization energies by replacing the less-accurate $E\x^{GGA}$ with more exchange. 

\ssec{Thermal DFT}
\label{sec:FTDFT}

DFT can be extended to finite temperatures by generalizing the Hohenberg-Kohn theorems and Kohn-Sham equations to thermal systems. Working within the grand canonical ensemble, Mermin generalized the HK theorems to equilibrium systems at finite temperatures constrained to fixed temperature and chemical potential\cite{M65}. The grand canonical potential of the system is written,
    \ben
        \hat{\Omega}=\hatH-\tau\hat{S}-\mu\hat{N}.
    \een
$\hatH$ is the electronic Hamiltonian, $\tau$ is the absolute temperature in Hartree units, $\mu$ is the chemical potential, $\hat{N}$ is the particle-number operator, and $\hat{S}$ is the entropy operator,
    \ben
        \hat{S} = -k_B \, \mathrm{ln} \, \hat{\Gamma}.
    \een
The statistical operator $\hat{\Gamma}$ is
    \ben \label{eq:statop}
        \hat{\Gamma} = \sum_{N,i} w_{\scriptscriptstyle\rm N,i} |\Psi_{\scriptscriptstyle\rm N,i}\rangle\langle \Psi_{\scriptscriptstyle\rm N,i}|,
    \een
where $|\Psi_{\scriptscriptstyle\rm N,i}\rangle$ are the orthonormal $N$-particle states and $w_{\scriptscriptstyle\rm N,i}$ are the normalized statistical weights satisfying $\sum_{\scriptscriptstyle\rm N,i}w_{\scriptscriptstyle\rm N,i}=1$. The statistical operator yields the thermally weighted, equilibrium density. The Mermin-Kohn-Sham (MKS) equations\cite{M65,KS65}, which resemble the ground-state KS equations, but with temperature-dependent eigenstates, eigenvalues, and effective potential, yield the MKS density,

    \ben
        n({\bf r})=\sum_i f_i\t |\phi\t_i(\bf r)|^2,
    \een
with $\phi_i^\tau({\bf r})$ equal to the $i^{\rm th}$ eigenstate and $f_i^\tau$ equal to the state's corresponding Fermi occupation. The MKS density is equal to the exact equilibrium density by definition.
To define the XC free energy, we decompose the universal functional and write the free energy as\cite{PPGB13},
    \begin{multline}
        A[n(\br)]=T\s[n(\br)]-\tau S\s[n(\br)]
        \\ +U\H[n(\br)]+A\t\xc[n(\br)]+V\ext[n(\br)].
    \end{multline}
$T\s$ is the non-interacting kinetic energy, $S\s$ is the non-interacting entropy, $U$ is the classical electrostatic repulsion, and $V\ext$ is the external potential\cite{FW71}. The XC free energy by definition is

    \begin{multline}
        A\t\xc[n(\br)]=\big(T[n(\br)]-T\s[n(\br)]\big)\\
        -\tau\big(S[n(\br)]-S\s[n(\br)]\big) \\
        +\big(V_{ee}[n(\br)]-U_H[n(\br)]\big),
    \end{multline}
where $T[n]$ and $S[n]$ are the interacting kinetic energy and entropy, and $V_{ee}[n]$ is the electron-electron interaction energy.

\sec{Results and Discussion}

\ssec{Finite-temperature two-legged construction}
A temperature-dependent TLA can be obtained from the FTAC. In this case, $b$ is defined as a temperature-dependent quantity and maintains the ZT $b$ parameter's dependence on the density through its definition in terms of X and XC density functionals:
    \ben
        b\t = \frac{A\xc\t - W\xc^{\tau,\lambda=1}}{A\x - W\xc^{\tau,\lambda=1}}.
    \een
Here, $b\t$ is defined in terms of free energies, which contain exchange and correlation entropy. Incidentally, $b$ at FT is not only a measure of the fraction of kinetic correlation to potential correlation; it is now comparing quantities that contain correlation entropy:
    \ben
    b\t = \frac{K\c\t}{|U\c\t|}.
    \een
Here, the numerator is the kentropic correlation, $T\c\t-\tau S\c\t$, which contains both kinetic and entropic contributions. The denominator is still the correlation potential energy, but it is now temperature dependent. In contrast to the ZT case, $b$ at FT is a measure of the fraction of correlation that is kentropic, though it is still true that this component of correlation is expected to dominate over $U\c\t$ in the high-density limit.

\ssec{Numerical Demonstrations}
The exact formulation of the FT TLA to the adiabatic connection curve serves as proof of principle of a FT extension to the original theory, by presenting two-legged curves that yield the exact $A\xc$ under the given density and temperature conditions.  It also provides a framework through which we can examine how the $b$ parameter depends on both density and temperature. Since $b$ indicates the curvature of the adiabatic connection integrand, it provides a measure of how much the exchange free energy and correlation free energy each vary under specific density-temperature conditions, as well as a window into how kentropic and potential correlation are balanced (or imbalanced) under certain conditions. 

To demonstrate the theory and begin such analysis in exactly solvable models, we first implement the TLA for the FT uniform electron gas. We then implement it for the FT asymmetric Hubbard dimer, to investigate how these relationships between limits depend on the strength of the model's on-site interaction, $U$. We reiterate here that, though the piecewise linear TLA curves are approximations to the adiabatic connection integrand, the integrated quantity of the $A\xc$ is given exactly by both the exact adibatic connection curve and the TLA curve in these model systems. Future work will build upon these theoretical foundation to apply this scheme to common density functional approximations at non-zero temperatures to generate entirely new FT hybrid approximations. 

\sssec{Uniform Electron Gas}

The uniform electron gas (UEG) is an important model system in physics which places an infinite system of equally spaced electrons in a "smeared-out" background of positive charges. Studies of the UEG have led to key realizations regarding superconductivity\cite{BCS57}, collective excitations from a quasiparticle perspective\cite{PB52,BP53}, and Fermi liquid theory\cite{GV08,BP08}. The UEG has also served as the basis of successful functional approximations to the XC energy, including the popular local density approximation,
    \ben
        E\xc^{LDA}[n] = \int d^3r \, n(\br)\epsilon\xc
        (n(\br)),
    \een
where $\epsilon_{xc}(n(\bf{r}))$ is the exchange-correlation energy per particle of a uniform electron gas with density $n$. The XC energy must reduce to the LDA when the density is uniform.
The UEG is useful for obtaining density-scaling relationships and coordinate-scaling conditions. For example, the exchange energy is the high-density limit,
    \ben
        E\x[n] = \lim_{\gamma\to\infty} \frac{E\xc[n_{\gamma}]}{\gamma},
    \een
where 
    \ben
    n_{\gamma}=\gamma^3 n (\gamma x, \gamma y, \gamma z).
    \een
The prefactor $\gamma^3$ is chosen to preserve normalization to $N$ electrons.

This work employs a parametrization of the XC free energy based on the uniform electron gas\cite{GDSM17} to obtain finite-temperature adiabatic connection curves. Few other highly-optimized parametrizations of the XC free energy currently exist.\cite{KSDT14,KDT18,MKH20}

A TLA is demonstrated for the FTAC of the UEG\cite{HMP22} in Figs.\ref{fig:rpt1t1000} - \ref{fig:r20tpt01} for different combinations of density and absolute temperature. A higher value of $r_s$, corresponding to a lower density, will result in the two legs meeting at a value closer to $\lambda = 0$ to accommodate the correlation free energy, which increases dramatically at lower interaction strengths for lower densities. The effect of temperature on $b\t$ is demonstrated in Figs. \ref{fig:r1comparison},  \ref{fig:b_vs_tau_1}, and \ref{fig:b_vs_tau_2}, showing its nonlinear behavior, distinct temperature-density regimes, and the switch in curve ordering for low and high temperatures at certain values of $r_s$. This highlights the complexity of the mixing parameter $b\t$, even for seemingly simple uniform systems. For FT systems, the identification of strong static or dynamic correlation demands combined assessment of density and temperature conditions. The $b\t$ parameter's capacity for greater flexibility in mixing X and C differently in future hybrid free energy approximations may be one way to capture the shifts in these ratios with a single functional form, while its demonstration in the exact solution of the FT UEG shows that such complexity is ubiquitous in electronic structure. 
       \begin{figure}
            \centering
            \includegraphics[width=\columnwidth]{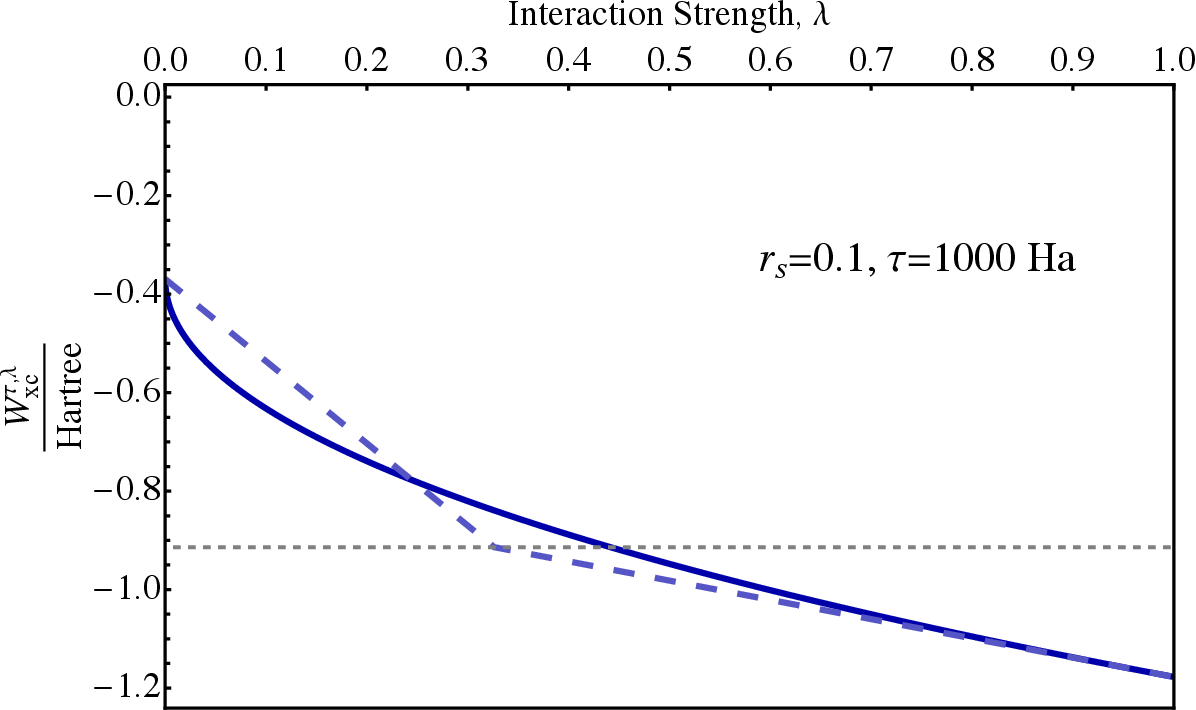}
            \caption{Comparison of the exact FTAC (solid) and the two-legged construction (dashed) for a Wigner-Seitz radius of 0.1 and an absolute temperature of 1000 Ha, as an example of a very high-density and very high-temperature UEG system.}
            \label{fig:rpt1t1000}
       \end{figure} 
       
       \begin{figure}
            \centering
            \includegraphics[width=\columnwidth]{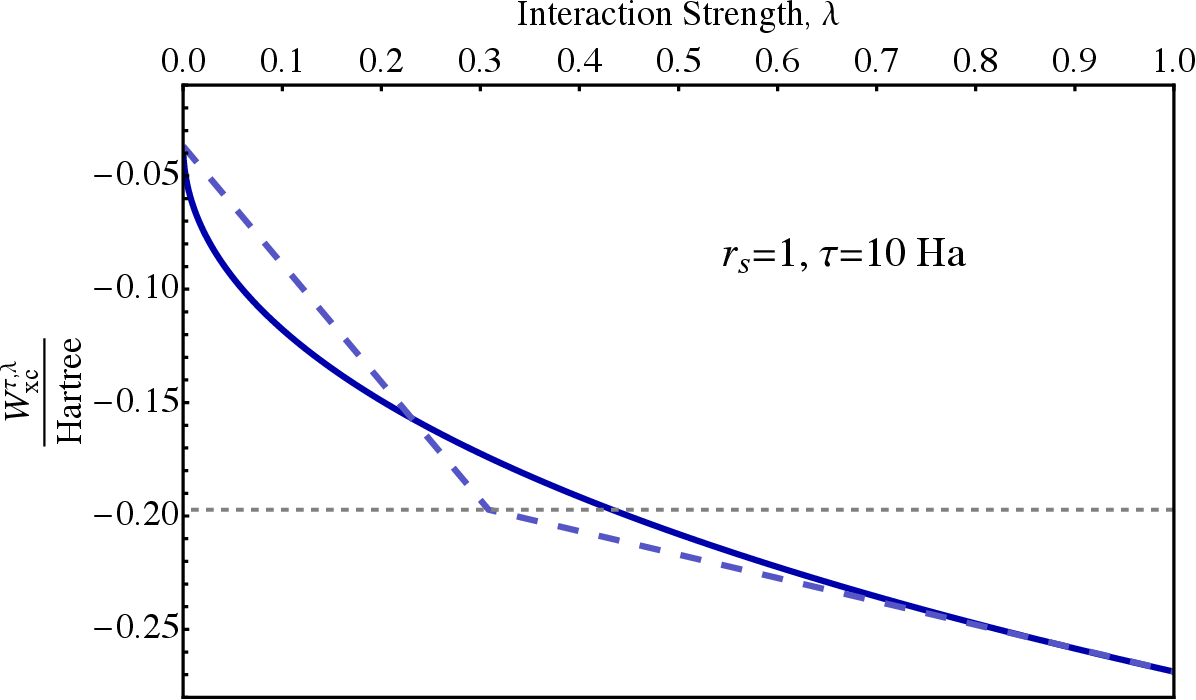}
            \caption{Comparison of the exact FTAC (solid) and the two-legged construction (dashed) for a Wigner-Seitz radius of 1 and an absolute temperature of 10 Ha, as an example of a high-density and high-temperature UEG system.}
            \label{fig:r1t10}
       \end{figure}
       
       \begin{figure}
            \centering
            \includegraphics[width=\columnwidth]{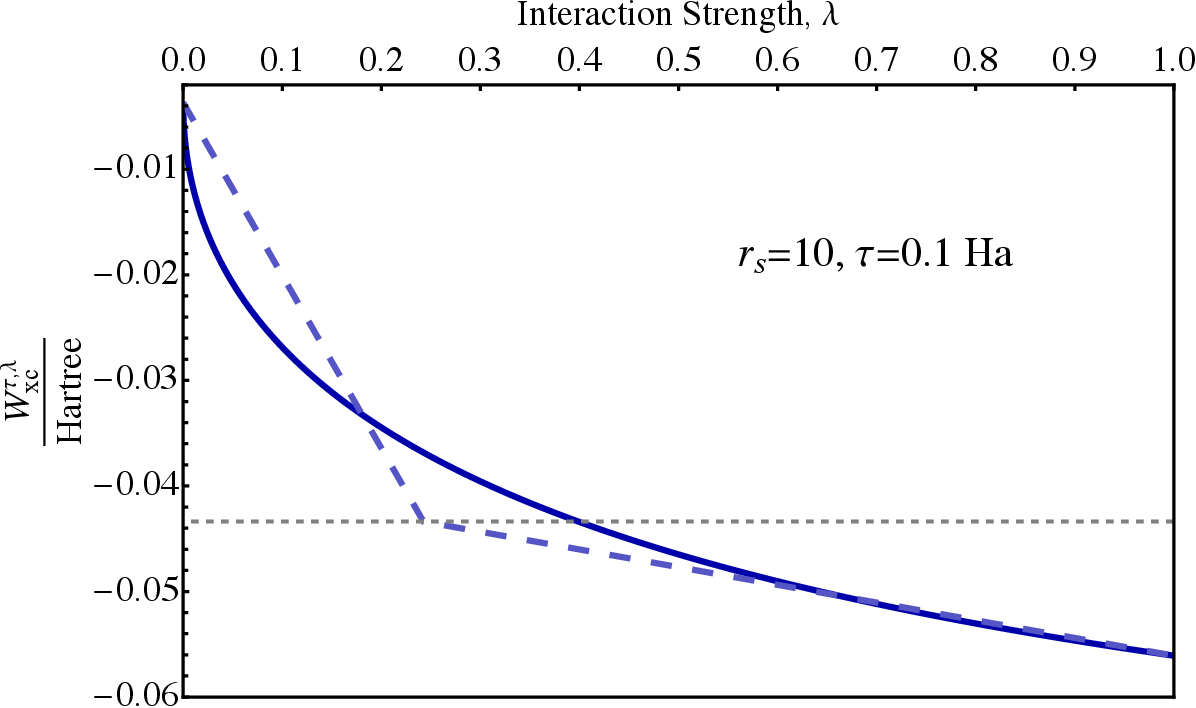}
            \caption{Comparison of the exact FTAC (solid) and the two-legged construction (dashed) for a Wigner-Seitz radius of 10 and an absolute temperature of 0.1 Ha, as an example of a low-density and low-temperature UEG system.}
            \label{fig:r10tpt1}
       \end{figure}

       \begin{figure}
            \centering
            \includegraphics[width=\columnwidth]{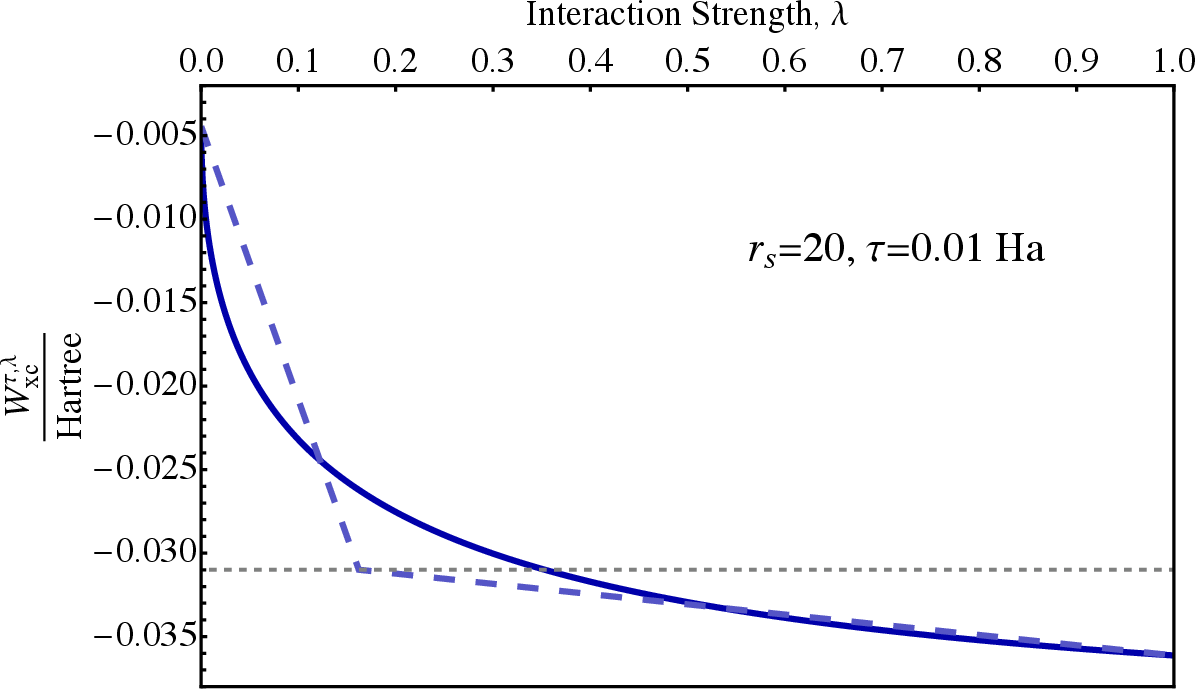}
            \caption{Comparison of the exact FTAC (solid) and the two-legged construction (dashed) for a Wigner-Seitz radius of 20 and an absolute temperature of 0.01 Ha, as an example of a very low-density and low-temperature UEG system.}
            \label{fig:r20tpt01}
       \end{figure}
       
       \begin{figure}
            \centering
            \includegraphics[width=\columnwidth]{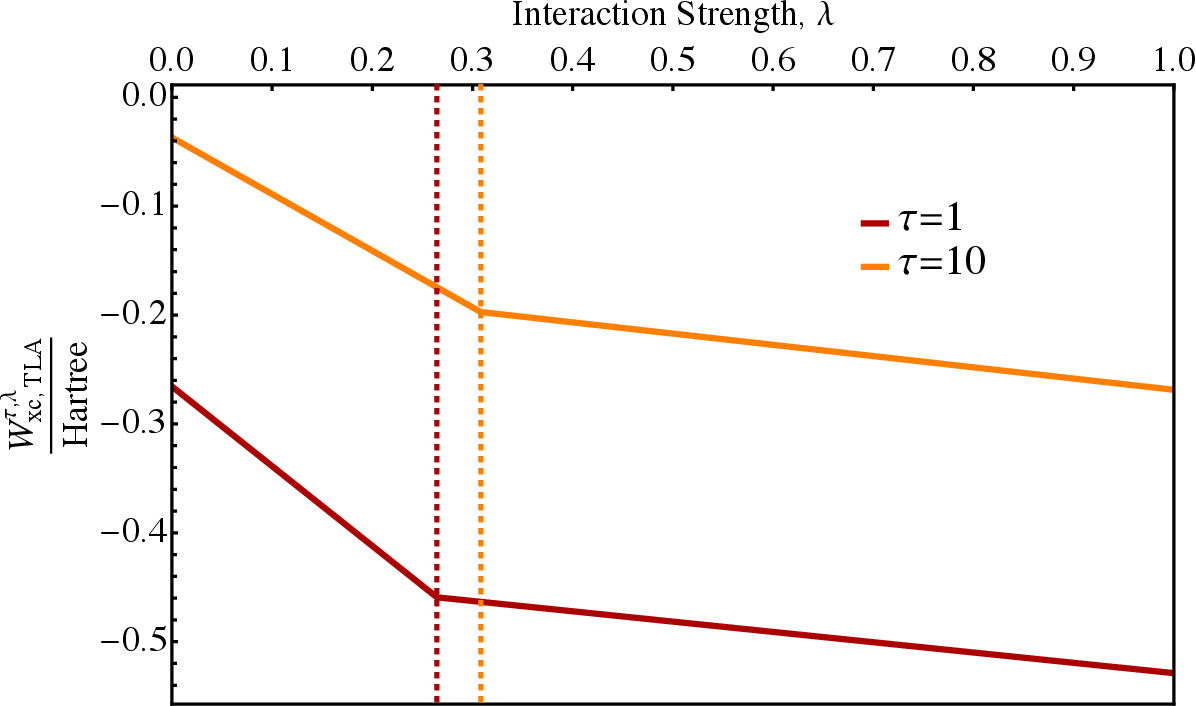}
            \caption{Two-legged construction of the finite-temperature adiabatic connection is plotted at two different temperatures for $r_s = 1$. Dashed lines represent the corresponding temperature-dependent b values.}
            \label{fig:r1comparison}
       \end{figure}

       \begin{figure}
            \centering
            \includegraphics[width=\columnwidth]{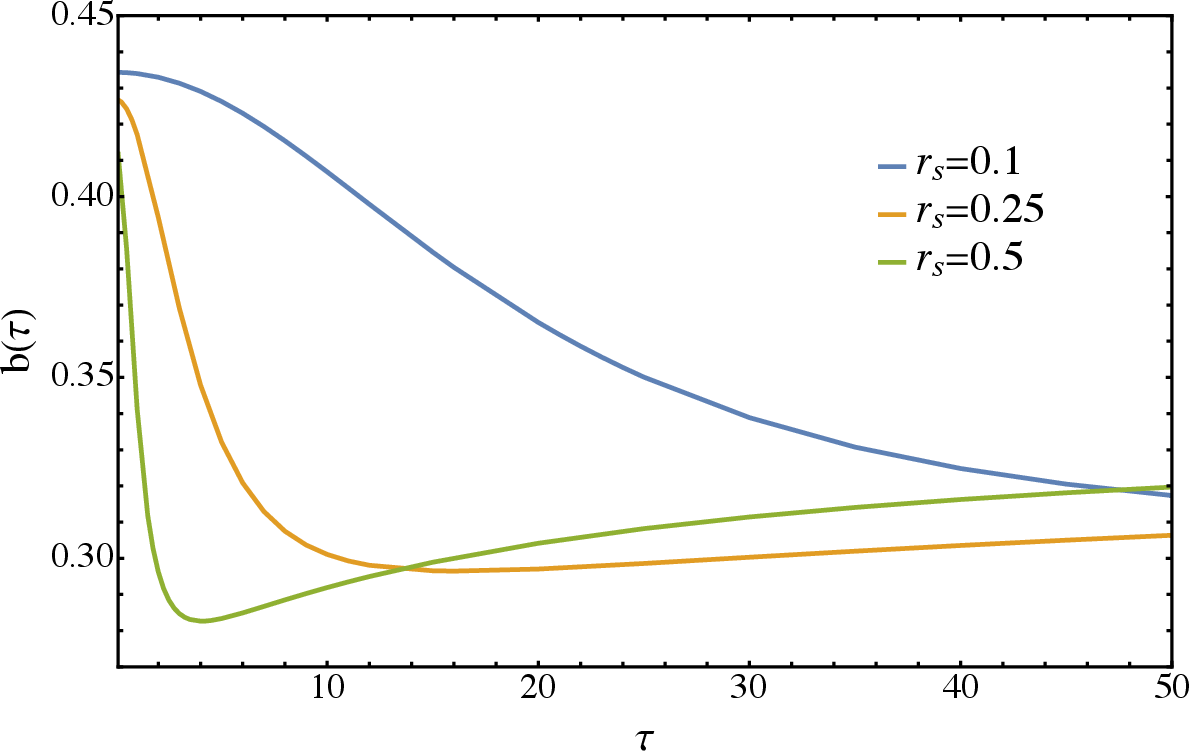}
            \caption{The point, $\lambda=b$, at which the two linear legs meet varies with temperature and density. The temperature dependence of $b$ is shown for three different densities.}
            \label{fig:b_vs_tau_1}
       \end{figure}

       \begin{figure}
            \centering
            \includegraphics[width=\columnwidth]{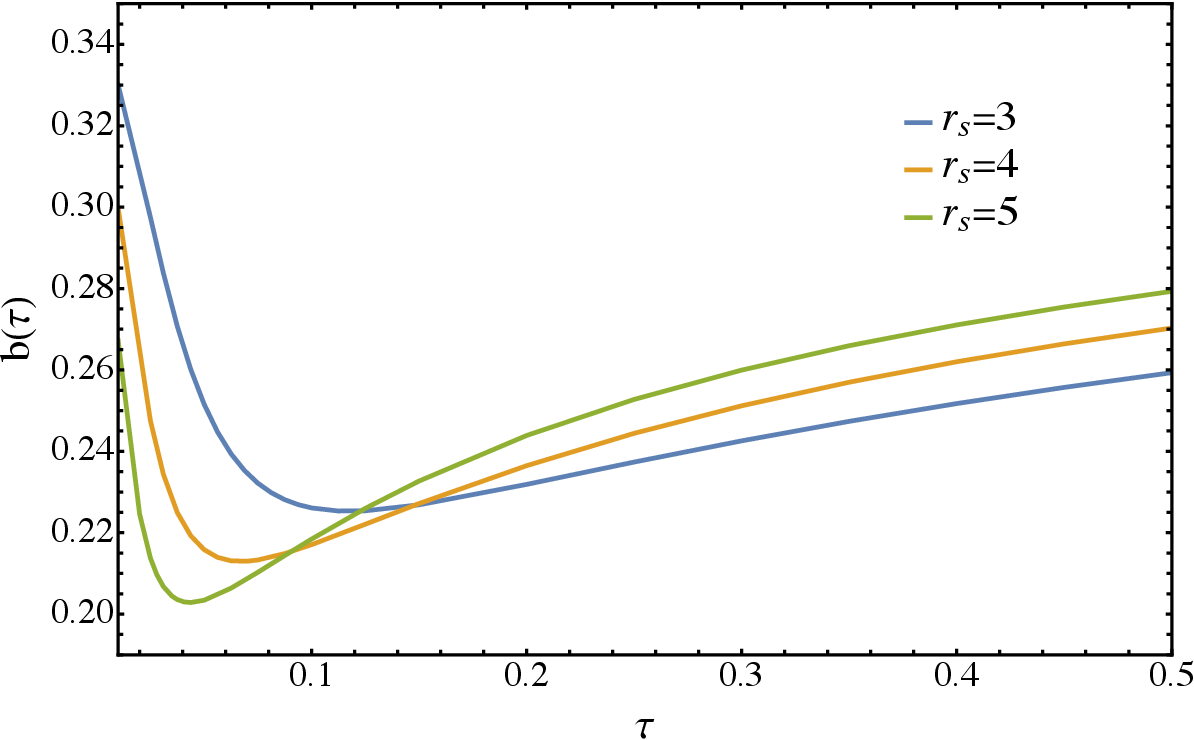}
            \caption{The point, $\lambda=b$, at which the two linear legs meet varies with temperature and density. The temperature dependence of $b$ is shown for three different densities, demonstrating swapped ordering with $r_s$ at low and high temperaturesfor these densities.}
            \label{fig:b_vs_tau_2}
       \end{figure} 

\sssec{Hubbard Dimer}

To go beyond the uniformity of the FT UEG, we next move to a demonstration using a non-uniform exactly solvable model with additional capabilities for modeling strongly correlated interactions. While density functional theory is exact in principle, density functional approximations often fail when correlations become strong. The Hubbard dimer\cite{CFSB15,SPB16} is a solvable, many-body, truncated two-site lattice model that is useful for investigating extreme limits of electronic structure systems, including those with non-uniform densities. As a lattice model, it is by nature well-suited for the treatment of strong correlation because the model allows for short-ranged interactions only. This seemingly simple truncated version accommodates up to four electrons, leading to a minimal basis set. The ground-state Hamiltonian for the Hubbard model is typically written as
    \ben
        \hat{H} = -t\sum_{\sigma}(\hat{c}^{\dagger}_{1\sigma}\hat{c}_{2\sigma} + \text{H.c.}) + U \sum_i( \hat{n}_{i\uparrow}\hat{n}_{i\downarrow}) + \sum_{i,\sigma} v_{i} \hat{n}_{i,\sigma}
    \een
where $t$ is the electron hopping strength between each site (analogous to the kinetic energy), $\hat{c}^{\dagger}_{1\sigma}$ and $\hat{c}_{2\sigma}$ are electron creation and annihilation operators, $\hat{n}_{i\sigma}$ is the number operator, $U$ is the (short-ranged) Coulomb repulsion experienced by two electrons occupying the same site, and $v_i$ is the external potential corresponding to each site.

The density of the system is controlled by the parameter $\Delta n = n_2 - n_1$, the occupation difference, which informs the potential difference $\Delta v$. The symmetric Hubbard model is obtained by setting $\Delta v = 0$, while the asymmetric model is obtained by setting $\Delta v \neq 0$. 

 Including temperature $\tau$, in this truncated model has shown to be useful in the analysis of finite-temperature DFT, and even for the generalization of existing theorems \cite{SSB16,SPB16,SB20}. The many-body solutions for the finite-temperature case can be found in the Appendix of Ref. \cite{SB20}. By controlling all of the parameters in this model system (\textit{i.e.} $U$, $t,\Delta v$, and $\tau$), we can reproduce the adiabatic connection curve by setting $U\to \lambda U$. We also chose to keep the chemical potential at $\mu= U/2$, for an average particle number of two for these calculations. Lastly, we note that, due to the truncated Hilbert space, we choose to limit the range of temperatures to $\tau <2  ~ a. u.$ While the UEG is useful for exploring uniform densities at FT, we can now use the Hubbard dimer to investigate non-uniform densities via the occupation difference $\Delta n \neq 0$. 

For the finite-temperature dimer, we first start by defining for an average particle number of $N=2$, the non-interacting density,  
\ben
\Delta n^{\tau}(\alpha,\phi) = -2\sin{\phi}\tanh{\alpha},
\een
where the density is dependent on three parameters $\alpha = (4\tau \cos{\phi})^{-1}$,  $\sin{\phi} = x/\sqrt{1+x^2}$ and $x = \Delta v_s/2t$. The Hartree energy is defined as $U_{H}={U}(1 + \frac{\Delta n^{2}}{4})$, and since we are observing the average $N=2$ case, the exchange energy is simply  $E_{_{X}}=-U_{H}/2$. 
To repeat the FT TLA for the dimer, we start by defining the 

\ben
U_{_{XC}}=V_{ee}^{\tau}-U_{H}, 
\een
\noindent where $V_{ee}^{\tau}$ corresponds to the many-body electron interaction. As mentioned earlier, if we integrate this term over the interaction strength $\lambda$, we find finite-temperature XC free energy has the form of

\ben
A_{_{XC}}^{\tau}= \int_{0}^{1}~d\lambda~\frac{U_{_{XC}}}{\lambda}.
\een

 \begin{figure}
            \centering
            \includegraphics[width=\columnwidth]{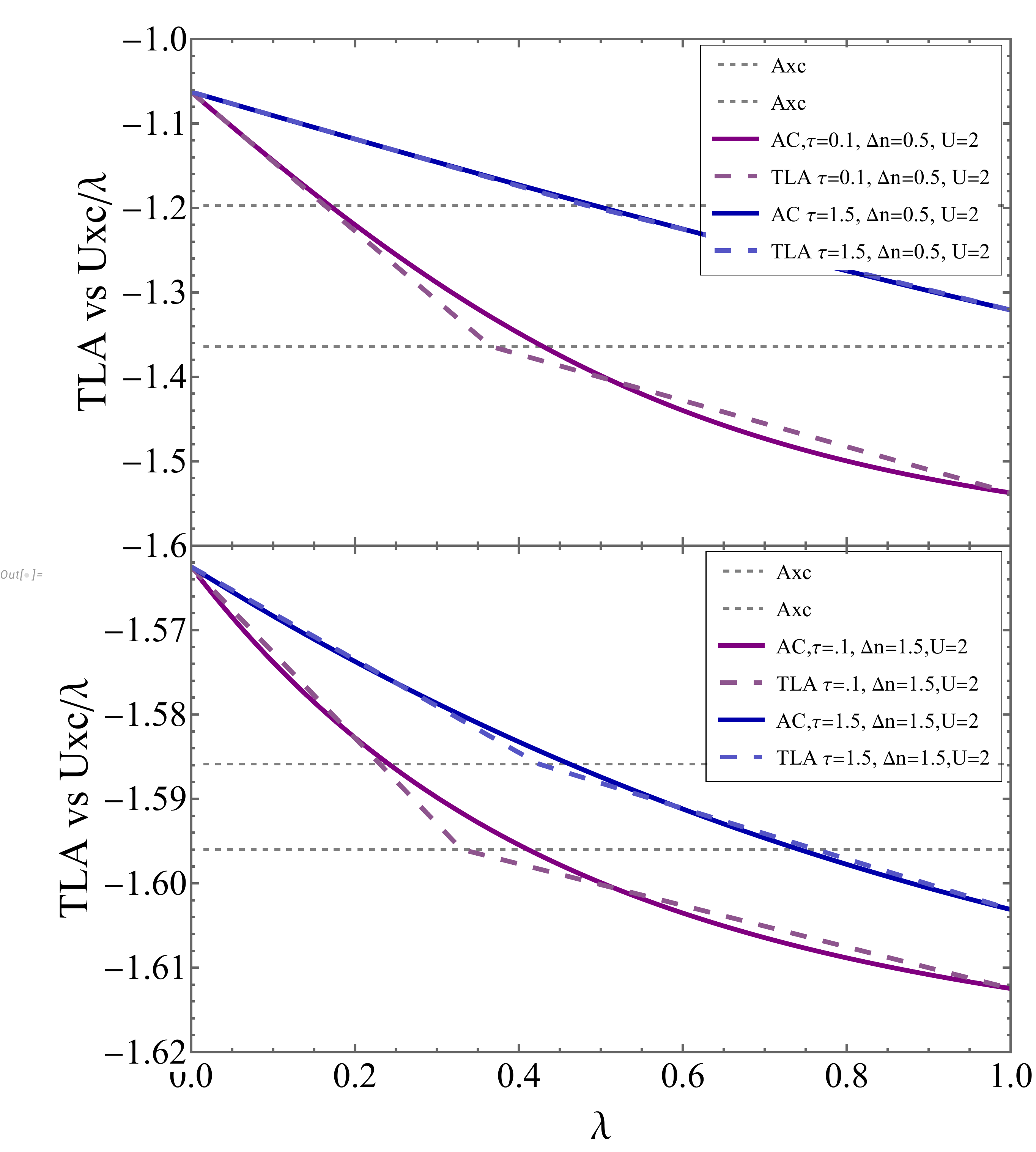}
            \caption{Solid lines correspond to the AC curve for varying densities ($\Delta n$) and temperatures ($\tau$) as a function of the coupling strength $\lambda$, for the finite temperature dimer for $U=2$.}
            \label{TLA_vs_Uxclambda_comb_highDen}
       \end{figure}

In Fig.~\ref{TLA_vs_Uxclambda_comb_highDen}, we compare the TLA approximation with the true AC at the same temperature, interaction, and density conditions. As expected, we find good agreement at the two extreme limits ($\lambda =0,1$), and deviation for $\lambda =b$, though the integrated $A\xc\t$ is the same by construction. Finally, in Figs.~\ref{fig:contour_deltan_U1} and ~\ref{fig:contours}, we plot contours of  $b\t$ as a function of temperature. We vary the density in Fig \ref{fig:contour_deltan_U1} while holding $U=1$, and we vary the interaction strength in Fig \ref{fig:contours} for various $\Delta n$ values. For the dimer, we find that at higher temperatures, there is better agreement between the AC and the TLA, both for the lower ($\sim \Delta n =0.5$) and higher densities ($\sim \Delta n=1.5$) examined. This shows that this combination of conditions sees the higher temperatures push the system toward a ``high-density-like" curvature and more influence from $T\c\t$.  

    \begin{figure}[htbp]
            \centering
            \includegraphics[width=\columnwidth]{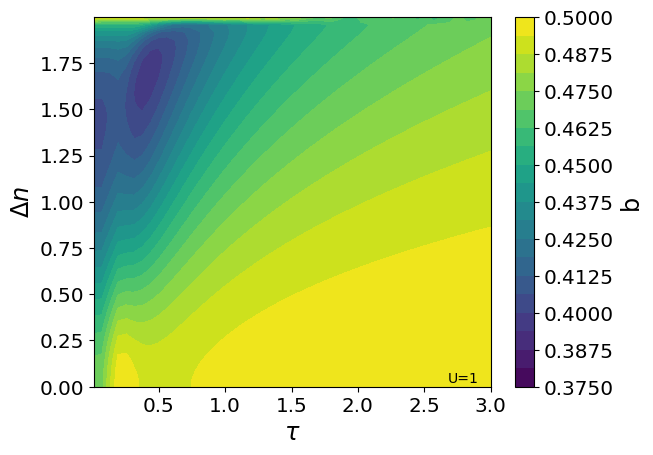}
            \caption{Contour of mixing parameter $b(\tau)$, plotted for onsite potential strength $U=1$, versus absolute temperature, $\tau$, and onsite density difference, $\Delta n$. }
            \label{fig:contour_deltan_U1}
       \end{figure} 

Looking at the contour plots of $b\t$, we find some interesting trends for this model system. First, for lower densities, the value of $b\t$ tends $\rightarrow 0.5$ even with increasing values of $U$. For higher densities, you must also decrease the values of $U$ in order for $b\t\rightarrow 0.5$, regardless of increasing temperature. Of particular note are the regions in both high- and low-density cases at low-to-intermediate temperatures, where nonlinear dependence on temperature, even at high $U$ values, indicates fluctuating ratios of dynamic and static correlation through the b values shifting balance of $T\c$ and $U\c$.

    \begin{figure*}[htbp]
            \centering
            \includegraphics[width=\textwidth]{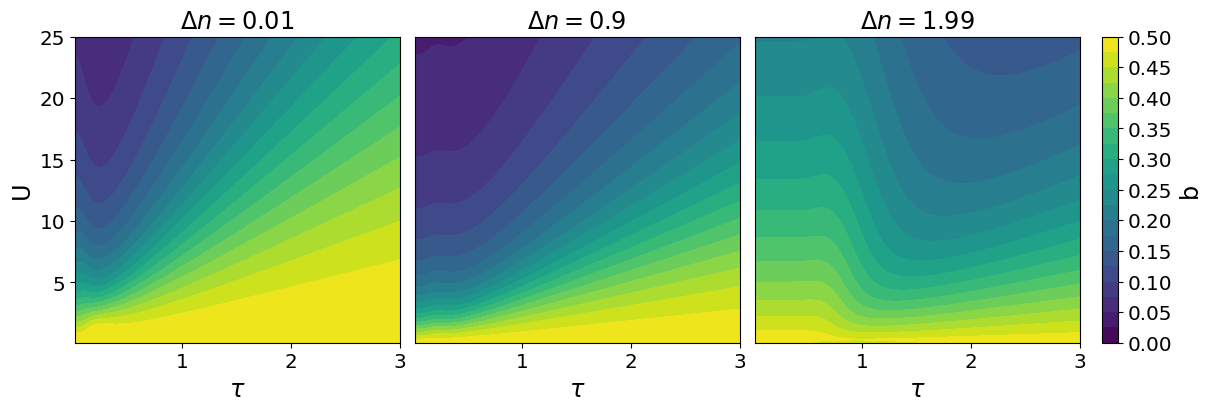}
            \caption{Contour of mixing parameter $b(\tau)$, plotted for various $\Delta n$ values versus absolute temperature, $\tau$, and onsite repulsion strength, $U$. For both small (left plot) and large (right plot) onsite density difference, nonlinear behavior is visible in low- to intermediate-temperature regimes and is dependent on $U$ value.}
            \label{fig:contours}
       \end{figure*} 

\sec{Conclusions}
We have presented the first approach to FT hybrid DFAs that produces a mixing parameter that is both density- and temperature-dependent.  This provides additional flexibility in its description of electron-electron interaction, which can be used to address the various conditions seen in WDM and other systems where fractional occupations and mixed-state descriptions are key. The FT TLA provides a straightforward pathway to create a new hybrid approximation from any existing FT GGA.

The definition of a temperature- and density-dependent $b$ parameter enables analysis of adiabatic connection integrand curvature with respect to temperature, density, and interaction strength, an object directly related a system's balance (or imbalance) of X and C free energies. This non-standard indicator of strong correlation character can be further broken down to compare the shifting magnitudes of kinetic, entropic, and potential components of the correlation free energy, all of which are generally elusive within standard analysis. Our use of the asymmetric Hubbard dimer shows clearly that, even with strong onsite repulsion between electrons, nonlinear dependence on temperature persists when the density is high enough, as one might see in WDM conditions of coupled high pressure and high temperature. Though these exact demonstrations, the $b\t$ value analysis clearly shows that determining whether a system exhibits strong static or dynamic correlation cannot be determined by a single condition (temperature or density) alone. Based on this, useful XC approximations for WDM are likely to need sophisticated approaches that adjust their handling of these different phenomena based on system conditions. 

Implementation and testing with FT GGAs\cite{KDT18,KPB23} is critical for addressing this need and initial steps are already underway. It has been shown previously\cite{MKH20} that the most widely available FT hybrid approximation suffers from the overestimation of band gaps seen in PBE0, its ZT limit. Implementing a FT TLA-based hybrid could alleviate some of this by better replicating true temperature dependence in the straightfoward form of a GGA hybrid. In analysis of $b$'s dependence on temperature, density, and reduced density gradient, we will evaluate whether these new, non-empirical hybrid parameters replicate the density scaling or other exact conditions seen in either meta GGAs or other approximation schemes at the hybrid ``rung" of DFAs. Futhermore, the parameter $b$ and the accuracies of the GGA FT TLA can be used as proxies for models of the finite-temperature exchange hole and other DFT objects. This numerical approach to questions about X and XC hole behavior is especially useful at non-zero temperatures, where useful mathematical tools, like the ZT sum rule, are known to break down due to particle number fluctuations in the grand canonical ensemble. 

A natural question is whether this can (or should) be applied to meta GGAs or other higher-order approximation schemes where one expects $E\xc$ to be more accurate than $E\x$. The dependence of $b$ on the kinetic energy density in the mGGA case would imply dependence on orbitals and so could result in an effect akin to the implicit semilocal form of "non-locality" seen in other mGGA properties. It is possible that $b$ depending on KED would mimic the tuning of KED dependence seen in ultra-non-local functional approximations, though it is unclear whether this would have any practical use or be only a tool for analysis of the underlying theory.

Two of the co-authors have ongoing work in progress on approximating the ACF to yield an LDA-like approximation to XC entropy,\cite{AHP25} which invites applying a similar TLA approach to these approximate curves capturing temperature dependence, instead of interaction dependence.  If a similar geometrically based rewriting of the integral using straight-line segments exists, it could highlight a novel method for approximating this complicated piece of XC free energy. It would also provide a key analysis of which parts of the XC correlation entropy (X or C pieces) are most dominant under certain temperature and pressure conditions.

The TLA approach can also be used in the context of the upside-down adiabatic connection,\cite{LB09,MGP25} whether at zero or non-zero temperatures.  In these cases, using the TLA with the exactly solvable asymmetric Hubbard dimer and UEG provide paths for approximating limiting values that are difficult to approach numerically in general cases. Finally, ongoing work on the FT generalized Kohn-Sham formalism\cite{HOP25} provides a rigorous framework for the TLA's resulting non-empirical hybrid approximations, as well as existing hybrid density functional approximations. These additional investigations highlight not only the practicality of a new hybrid approximation with a special type of explicit temperature dependence, but also how the FT TLA $b$ parameter yields a unique analytical tool for thermal DFT and beyond.

\sec{Acknowledgments} 
Fruitful discussions with Carsten Ullrich are gratefully acknowledged. 
B.H. and A.P.J's work is supported by the U.S. Department of Energy, National Nuclear Security Administration, Minority Serving Institution Partnership Program, under Awards DE-NA0003866 and DENA0003984. V.M.'s work is supported by Cottrell Scholar Award no. 28281, sponsored by Research Corporation for Science Advancement. We acknowledge all indigenous peoples local to the site of University of California, Merced, including
the Yokuts and Miwuk, and thank them for allowing us
to live, work and learn on their traditional homeland (see
\url{https://www.hypugaea.com/acknowledgments}).

\bibliographystyle{unsrt}
\bibliography{thermal,WDM,Master_TLAedit}

\end{document}